\documentclass[%
 reprint,
 amsmath,amssymb,
 aps,
]{revtex4-1}

\usepackage{subfigure}
\usepackage{graphicx}
\usepackage{dcolumn}
\usepackage{bm}
\usepackage[mathlines]{lineno}

\begin{document}
\preprint{APS/123-QED}

\title{Benford's Law in Astronomy}

\author{Theodoros Alexopoulos}
 \email{Theodoros.Alexopoulos@cern.ch}
\affiliation{%
 National Technical University of Athens
}%

\author{Stefanos Leontsinis}%
 \email{Stefanos.Leontsinis@cern.ch}
\affiliation{%
 National Technical University of Athens\\ Brookhaven National Laboratory
}%

\date{\today}

\begin{abstract}
Benford's law predicts the occurrence of the $n^{\mathrm{th}}$ digit of numbers in datasets originating from various sources all over the world, ranging from financial data to atomic spectra. It is intriguing that although many features of Benford's law have been proven, it is still not fully mathematically understood. In this paper we investigate the distances of galaxies and stars by comparing the first, second and third significant digit probabilities with Benford's predictions. It is found that the distances of galaxies follow the first digit law reasonable well, and that the star distances agree very well with the first, second and third significant digit.
\end{abstract}

\pacs{Valid PACS appear here}
\maketitle


\section{Introduction}
\label{sec:intro}
In $1881$ the astronomer and mathematician S. Newcomb made a remarkable observation with respect to logarithmic books \cite{Newcomb}. He noticed that the first pages were more worn out than the last. This led him to the conclusion that the significant digits of various physical datasets are not distributed with equal probability but the smaller significant digits are favored. In $1938$ F. Benford continued this study and he derived the law of the anomalous numbers \cite{benford_original}.

The general significant digit law \cite{digit_formula} for all $k\in\mathbb{N}$, $d_1\in{\{1,2,\ldots,9\}}$ and $d_k\in{\{0,1,\ldots,9\}}$, for $k\geq 2$ is

\begin{equation}
\label{formula_all_digits}
P(d_1, d_2, \ldots, d_k)=\log_{10}\left[1+\left(\sum_{i=1}^{k}d_i\times 10^{k-i}\right)^{-1}\right]
\end{equation}
where $d_k$ is the $k^{\mathrm{th}}$ leftmost digit. For example, the probability to find a number whose first leftmost digit is $2$, second digit is $3$ and third is $5$ is $P(d_1=2,d_2=3,d_3=5)=\log_{10}(1+1/235)=0.18\,\%$.

For the first significant digit can be written as

\begin{equation}
\label{formula_first_digit}
P(k)=\log_{10}\left(1+\frac{1}{k}\right),\ k=1,2,\ldots, 9
\end{equation}

This law has been tested against various datasets ranging from statistics \cite{non_physics_cite1} to geophysical sciences \cite{non_physics_cite2} and from financial data \cite{non_physics_cite3} to multiple choice exams \cite{non_physics_cite4}. Studies were also performed in physical data like complex atomic spectra \cite{complex_atomic_spectra}, full width of hadrons \cite{hadron_width} and half life times for alpha and $\beta$ decays \cite{alpha_decays, beta_decays}. 

An interesting property of this law is that it is invariant under the choice of units of the dataset (scale invariance) \cite{scale_invariant}. For example, if the dataset contains lengths, the probability of the first significant digits is invariant in the case that the units are chosen to be meters, feet or miles.

Still, Benford's law is not fully understood mathematically. A great step was done with the extension of scale to base invariance (the dependance of the base in which numbers are written) by Theodore Hill \cite{Hill_base}. Combining these features and realising that all the datasets that follow Benford's law are a mixture from different distributions, he made the most complete explanation of the law. Another approach in the explanation of the logarithmic law was examined by Jeff Boyle \cite{Jeff_Boyle} using the Fourier series method.

\section{New Numerical Sequences and Benford's Law}

A simple example of Benford's law is performed on numerical sequences. It is already proven that the Fibonacci and Lucas numbers obey the Benford's law \cite{ref:fibonacci}. The three additional numerical sequences considered in this paper are:
\begin{itemize}
\item Jacobsthal numbers ($J_n$), defined as 
	\begin{itemize}
		\item $J_0=0$
		\item $J_1=1$
		\item $J_n=J_{n-1}+2J_{n-2},\ \forall\ n>1$
	\end{itemize}
\item Jacobsthal-Lucas numbers ($JL_n$), defined as 
	\begin{itemize}
		\item $JL_0=2$
		\item $JL_1=1$
		\item $JL_n=JL_{n-1}+2JL_{n-2},\ \forall\ n>1$
	\end{itemize}
\item and Bernoulli numbers ($B_n$), defined by the contour interval
	\begin{itemize}
		\item $B_0=1$
		\item $B_n=\frac{n!}{2\pi i}\oint \frac{z}{e^z-1} \frac{dz}{z^{n+1}}$
	\end{itemize}
\end{itemize}

\begin{figure}[h] \centering
        \subfigure[] {
                \includegraphics[scale=0.27]{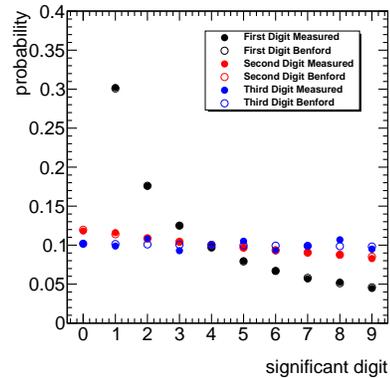}
                \label{fig:jacobsthal}
        }
        \subfigure[] {
                \includegraphics[scale=0.27]{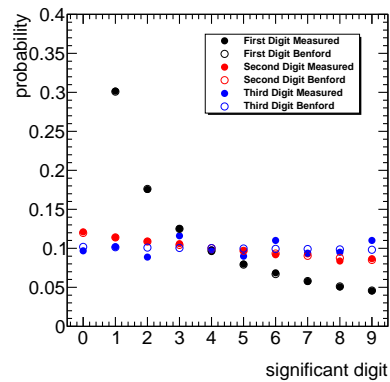}
                \label{fig:jacobsthallucas}
        }
        \subfigure[] {
                \includegraphics[scale=0.27]{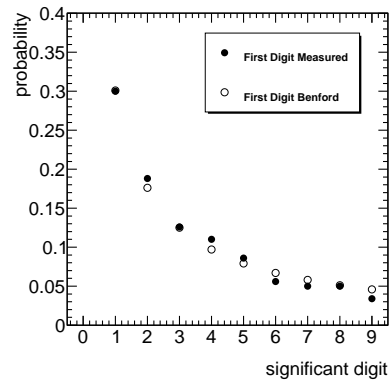}
                \label{fig:bernulli}
        }
        \caption{Comparison of Benford's law (empty circles) predictions and the distribution of the first, second and third significant digit of the (a) Jacobsthal, (b) Jacobsthal-Lucas and (c) Bernoulli sequences (full circles). The probabilities for the first digit is plotted with black, the second with red and the third with blue circles according to Benford's law.}
        \label{fig:numerical}
\end{figure}

A sample of the first $1000$ numbers of these sequences is used to extract the probabilities of the first significant digit to take the values $1,2,...,9$ and the second and third significant digits to be $0,1,...,9$. The results can be seen in figure \ref{fig:numerical}. Full circles represent the result from the analysis of the Jacobsthal and Jacobsthal-Lucas numbers and the empty circles indicate the probabilities calculated from Benford's formula (equation \ref{formula_all_digits}). It is clear that all three sequences follow Benford's law for the first (black), second (red) and third (blue) significant digit. 

In the following sections we examine the distances of stars and galaxies and compare the probabilities of occurrence of the first, second and third significant digit with Benford's logarithmic law. If the location of the galaxies in our universe and the stars in our galaxy are caused by uncorrelated random processes, Benford's law might not be followed because each digit would be equiprobable to appear. To our knowledge this is the first paper that attempts to correlate cosmological observables with Benford's law.

\section{Applications to Astronomy}
Cosmological data with accurate measurements of celestial objects are available since the 1970s. We examine if the frequencies of occurrence of the first digits of the distances of galaxies and stars follow Benford's law.

\subsection{Galaxies}
We use the measured distances of the galaxies from references \cite{data_galaxies_01,data_galaxies_72}. The distances considered on this dataset are based on measurements from type II Supernova and all the units are chosen to be $\mathrm{Mpc}$. The type-II supernova (SNII) radio standard candle is based on the maximum absolute radio magnitude reached by these explosions, which is $5.5\,\times 10^{23}\,\mathrm{ergs/s/Hz}$. 

\begin{figure}[h] \centering
        \includegraphics[scale=0.21]{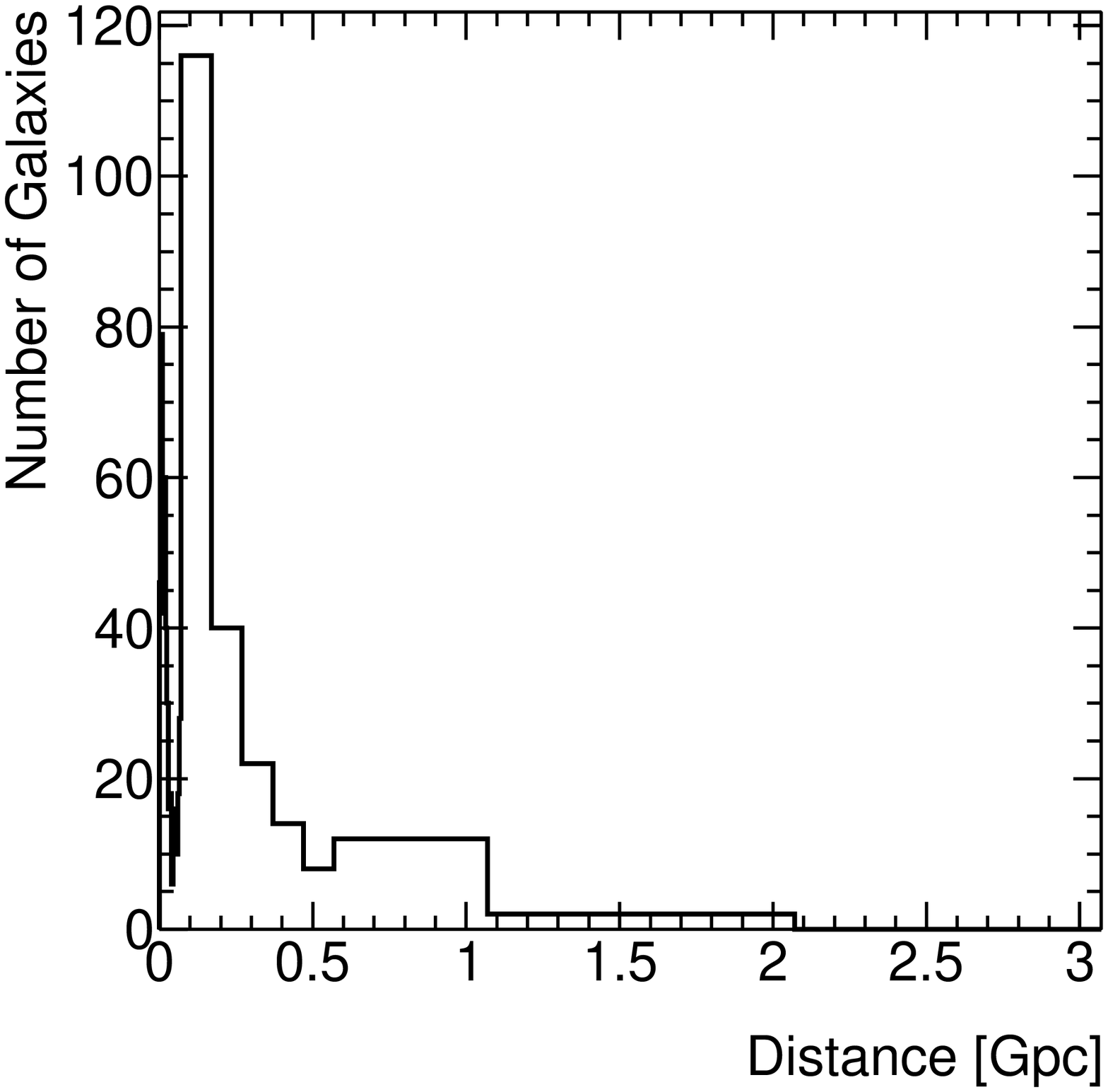}
        \includegraphics[scale=0.21]{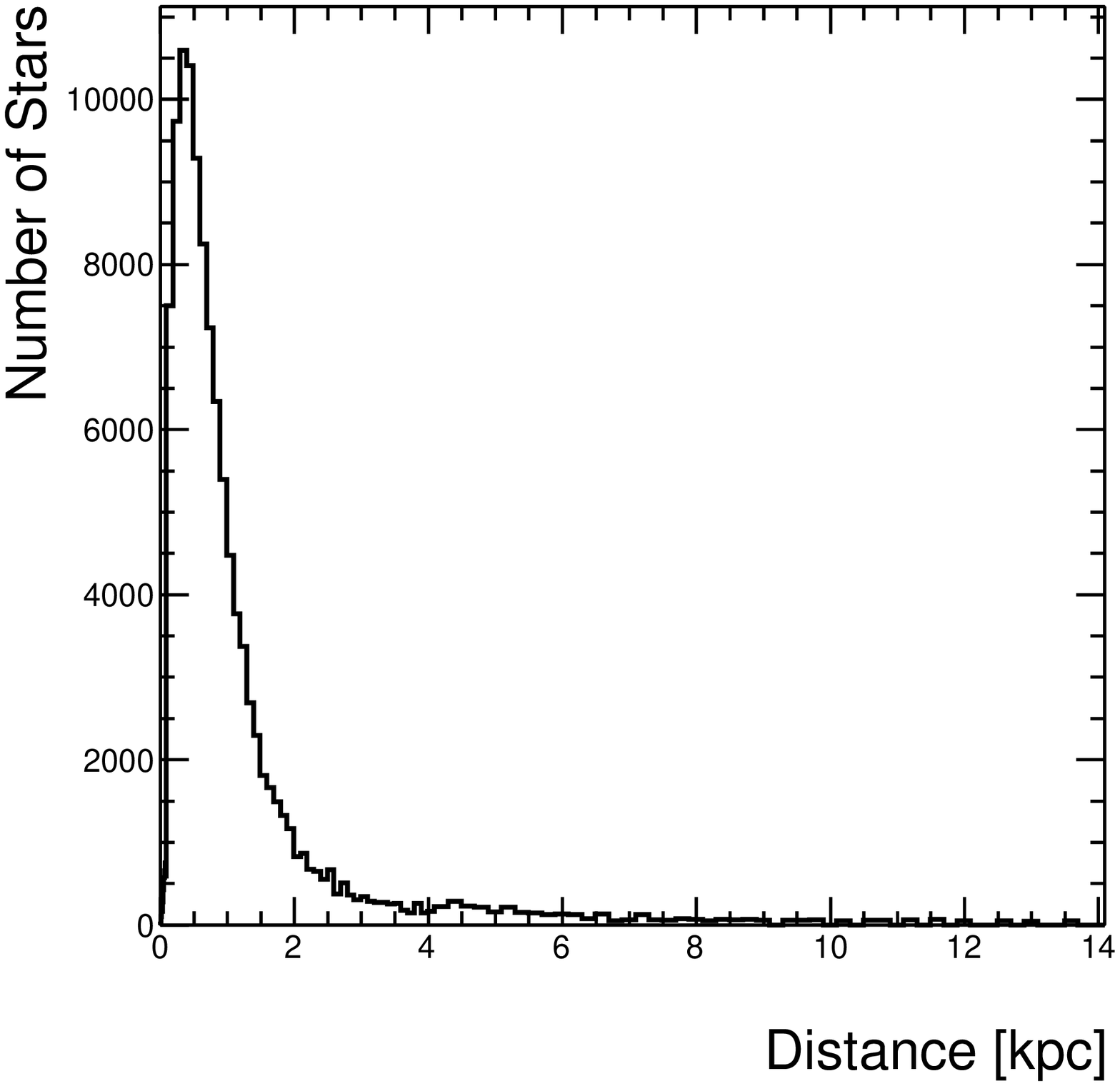}
        \label{fig:dataset}
        \caption{Complete dataset from where the measurements for the galaxies (left) and stars (right) is shown.}
\end{figure}

The total number of galaxies selected is 702 with distances reaching $1660\,\mathrm{Mpc}$ (see figure \ref{fig:dataset} left). The results can be seen in figure \ref{fig:galaxies} where with open circles we notate Benford's law predictions and the  measurements with the circle. Unfortunately due to lack of statistics the second and the third significant digit cannot be analyzed. The trend of the distribution tends to follow Benford's prediction reasonably well.

\subsection{Stars}
\begin{figure}[h] \centering
        \subfigure[] {
                \includegraphics[scale=0.27]{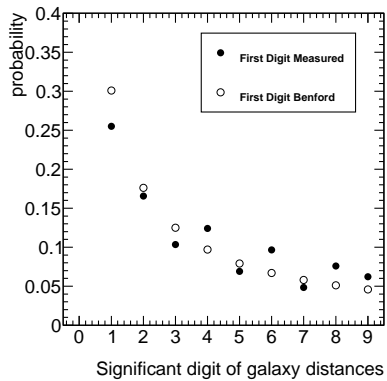}
                \label{fig:galaxies}
        }
        \subfigure[] {
                \includegraphics[scale=0.27]{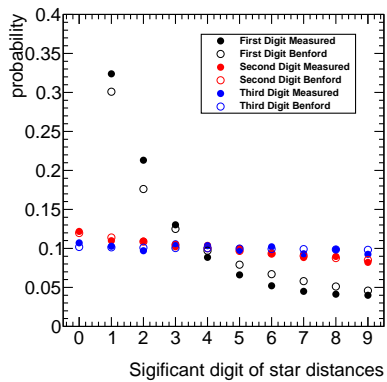}
                \label{fig:stars}
        }
        \caption{Comparisons of Benford's law (empty circles) and the distribution of the first (black), second (red) and third (blue) significant digit of the distances of the (a) galaxies and (b) stars (full circles).}
        \label{fig:stars_and_galaxies}
\end{figure}

The information for the distances of the stars is taken from the HYG database \cite{data_stars}. In this list $115\,256$ stars are included, with distances reaching up to $14\,\mathrm{kpc}$. The full dataset used for the extraction of the result can be seen in figure \ref{fig:dataset}. The result after analysing this dataset can be seen in figure \ref{fig:stars}. The first (black full circles) and especially the second (red full circles) and the third (blue full circles) significant digits follow well the probabilities predicted by Benford's law (empty circles).

\section{Summary}
The Benford law of significant digits was applied for the first time to astronomical measurements.  It is shown that the stellar distances in the HYG database follow this law quite well for the first, second and third significant digits.  Also, the probabilities of the first significant digit of galactic distances using the Type II supernova photosphere method is in good agreement with the Benford distribution; however, the errors are sufficiently large so that additional digits cannot be analyzed.  We note, however, that the plots in figure \ref{fig:dataset} indicate that selection effects due to the magnitude limits of both samples may be responsible for this behaviour and so it is not firmly established.  Therefore it is necessary to repeat this study using different galactic distance measures and larger catalogs of both galaxies and stars to see if the Benford law is still followed when larger distances are probed.  Such larger samples of galaxies would also allow the examination of second and perhaps third significant digits.

\acknowledgments
We would like to thank I.P. Karananas for the lengthy discussions on this subject. We would like also to thank Emeritus Professor Anastasios Filippas, the editor of JOAA and the reviewer for the valuable comments and suggestions.

The present work was co-funded by the European Union (European Social Fund ESF) and Greek national funds through the Operational Program "Education and Lifelong Learning" of the National Strategic Reference Framework (NSRF) 2007-1013. ARISTEIA-1893-ATLAS MICROMEGAS.

\end{document}